\documentclass[%
 reprint,
 amsmath,amssymb,
 aps,
]{revtex4-2}

\usepackage{graphicx}
\usepackage{dcolumn}
\usepackage{bm}
\usepackage{braket}
\usepackage{caption}
\usepackage{subcaption}
\usepackage{float}
\usepackage{amsmath}

\begin{document}

\preprint{APS/123-QED}

\title{Cooperative two-photon lasing in two Quantum Dots embedded inside Photonic microcavity}

\author{Lavakumar Addepalli}
\author{P.K. Pathak}
\affiliation{
 School of Physical Sciences, Indian Institute of Technology Mandi, Kamand, H.P., 175005, India}

\date{\today}

\begin{abstract}
We propose cooperative two-photon lasing in two quantum dots coupled to a single mode photonic crystal cavity. We consider both quantum dots are driven either incoherently or coherently using external pump. We incorporate exciton-phonon coupling using polaron transformed master equation. Using quantum laser theory, single and multi-photon excess emission (difference between emission and absorption) into cavity mode are investigated. The single and two-photon excess emission contribute to cavity photons, predominantly. Varying the pump strength can lead to single-photon excess emission change from negative to positive and thus by appropriately selecting pump strength single-photon excess emission can be made negligible. 
\end{abstract}

\maketitle


\section{\label{sec:level1}Introduction}
Cooperative effect in the ensemble of $N$ two level emitters interacting with a common electromagnetic field leads to superradiant emission\cite{Dicke1954,gross1982}, where emitted power is proportional to $N^2$. Superradiance occurs as a result of emission from correlated atomic dipoles\cite{temnov2009,auffeves2011,mascarenhas2013}. The quantum correlation is established through photon exchange between emitters. Superradiance has been observed in various quantum systems such as trapped atoms\cite{goban2015,devoe1996}, superconducting qubits\cite{Lambert2016} including quantum dots\cite{scheibner2007,kim2018}. In superradiant laser, steady state superradiance has been achieved using incoherent pump such that the correlation between emitters is maintained. Therefore, incoherently pumped ensembles of atoms inside a cavity under weak coupling regime leads to continuous collective emission of coherent radiation. In this paper we explore the question what will happen if the cavity is coupled under strong coupling regime. We particularly discuss single-photon and two-photon lasing in cooperative emission.
On the other hand, coherent pump dresses up the quantum emitter states in a way that alters their behavior. The modification leads to some fascinating effects, like changing absorption and emission properties of the system. For finite detuning between pump and emitter states, the population inversion is achieved between these dressed state. A dressed state laser has been realized by placing ensemble of such dressed emitters in a high quality cavity that is resonant with the transition between the dressed states\cite{Zhu1987,Boone1989,Boone1990,Davidovich1987}. Further, two-photon gain in dressed-state lasers has been realized by appropriately tuning the cavity frequency to the transitions between the dressed states satisfying two-photon emission condition\cite{Lewenstein1990,Law1991,Zakrzewski1991,Gauthier1992}. 
 Here, we also consider coherently driven two quantum dots (QDs) embedded inside a single mode photonic microcavity. The cooperative two-photon lasing is different than two-photon dressed state laser where emitters independently emit photons in cavity mode.

 Semiconductor QDs are artificial atoms due to strong confinement of electron hole pairs in all three directions, leading to discrete energy levels. The systems having QDs coupled to photonic crystal microcavity\cite{majumdar2012} have emerged as a potential candidates for realizing on-chip cavity quantum electrodynamics (CQED)\cite{englund2007,snijders2018,flagg2009,lagoudakis2017,pathak2011,stevenson2006}. This has also paved the way for integrated photonic networks with applications in opto-electronics\cite{Zhou2016}, quantum information processing\cite{Imamoglu1999,Kiraz2004}, quantum metrology\cite{Vittorio2006}, and realization of single photon source\cite{Pascale2017,pelton2002,yao2010}.
 However, unlike in the atomic systems, exciton-phonon interactions are inevitable in QDs that lead to strong dephasing, and the phonon-mediated cavity mode feeding predominantly. Various other phenomenon such as population inversion in two-level QDs has also been realised due to exciton-phonon interactions. Thereby, the exciton-phonon coupling at low temperatures, is essential to be included in the dynamics of QD-CQED systems\cite{nazir2016,roy2011}.
 
 Recently, highly efficient single QD laser has been realized in weak coupling regime\cite{Reitzenstein2008,Strauf2006,Nomura2009} using incoherent pump, where charge carriers are created in wetting layer and then diffuse into the QD or charge carriers are injected electrically into the QD. Further, in strongly coupled QD-cavity system,  coexistence of Rabi oscillations and lasing has been demonstrated\cite{Nomura2010laser, Elena2009}. Here, we predict cooperative two-photon lasing in two-QD coupled with a single-mode photonic crystal cavity and the QDs are driven incoherently or coherently using an external pump. We include exciton-phonon interactions non-perturbatively by making polaron transformation \cite{Mahan1990,xu2016} and derive the master equation using Born-Markov approximation\cite{carmichael1999}. Thereafter, we obtain the steady-state populations and mean cavity photon number using quantum optics toolbox\cite{tan1999}. We have also derived a simplified master equation (SME), which provides a clear view of various phonon-induced scattering processes in the system. Using SME, we have calculated the single-photon, multi-photon emission and absorption rates in steady state exactly. Lasing action in the system is observed when the emission rate exceeds absorption and other losses. Therefore, we present ``Excessive Photon Emission" in the cavity mode, i.e., the difference between emission and absorption in cavity mode. We calculate excessive photon emission separately for single-photon and multi-photon processes.

The paper is organised as follows. In section II we discussed the lasing in the incoherently pumped two-QD cavity system and in section III, coherently pumped system where we have calculated steady-state populations, cavity photon statistics, emission and absorption rates. We have also derived the simplified master equation treating the QD exciton-phonon coupling by making a polaron transformation and using the Born-Markov approximation and conclusions are made in section IV. 

\section{\label{sec:level3}Cooperative two-photon lasing in incoherently pumped two-QDs}
\begin{figure*}
    \centering
    \includegraphics[width=\textwidth]{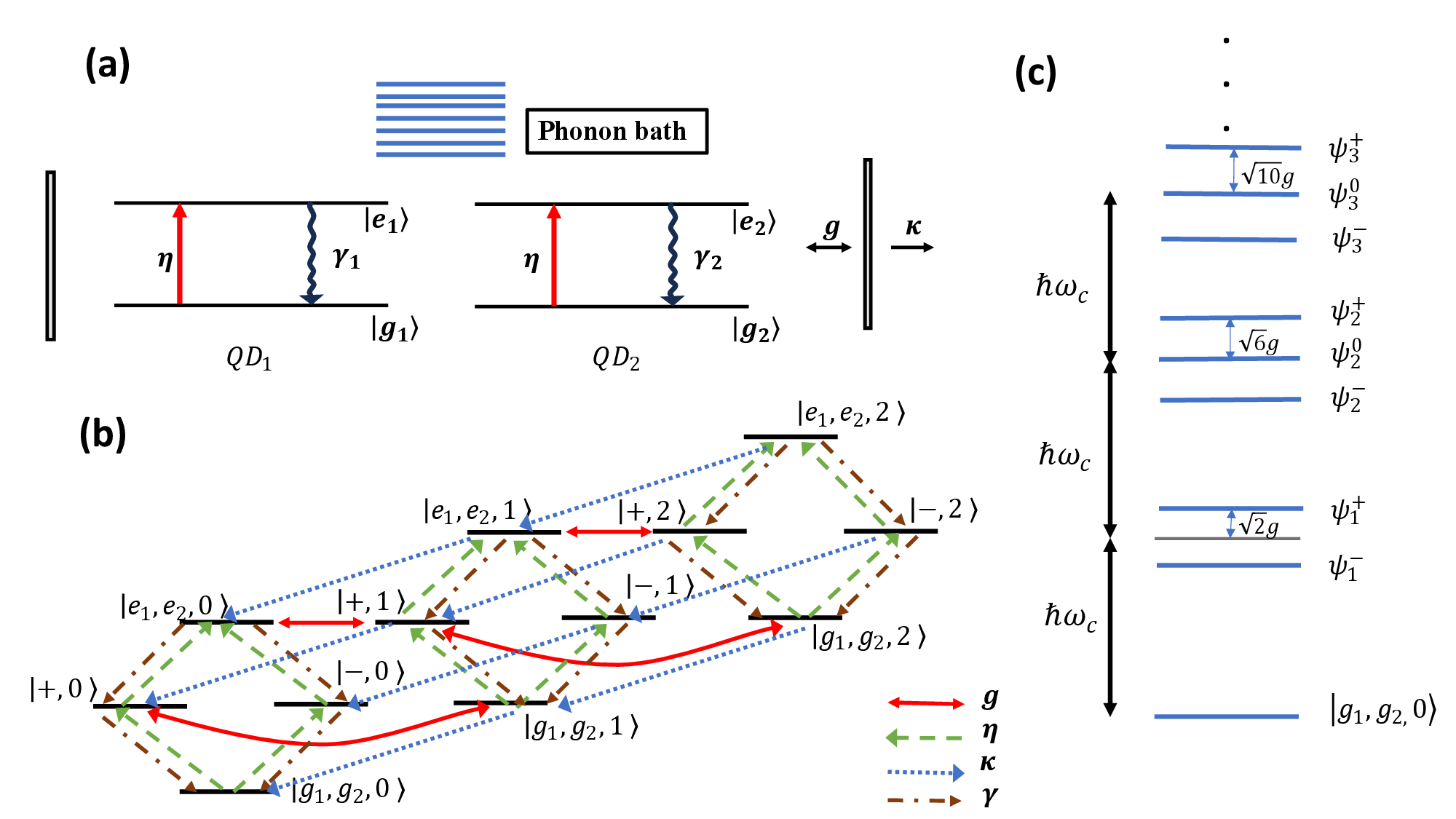}
    \caption{a)Schematic for energy levels of two-QD embedded in single mode microcavity with external pumping. b) Transitions between collective QDs-cavity states induced by cavity coupling, incoherent pumping, cavity decay and spontaneous emission are shown (Phonon induced processed are not labelled). c) The cavity dressed two-QD states for $\Delta=0$. The dressed state are given by, $\psi_1^\pm=\frac{\ket{+,0}\pm\ket{g_1,g_2,1}}{\sqrt{2}}$, $\psi_2^0=\frac{-\sqrt{2}\ket{e_1,e_2,0}+\ket{g_1,g_2,2}}{\sqrt{3}}$, $\psi_2^\pm=\frac{\ket{e_1,e_2,0}\pm\sqrt{3}\ket{+,1}+\sqrt{2}\ket{g_1,g_2,2}}{\sqrt{6}}$, $\psi_3^0=\frac{-\sqrt{3}\ket{e_1,e_2,1}+\sqrt{2}\ket{g_1,g_2,3}}{\sqrt{5}}$, $\psi_3^\pm=\frac{\sqrt{2}\ket{e_1,e_2,1}\pm\sqrt{5}\ket{+,2}+\sqrt{3}\ket{g_1,g_2,3}}{\sqrt{10}}$.}
    \label{fig:Fig1}
\end{figure*}

We consider two separate QDs coupled to a single mode photonic-crystal cavity. In Fig. \ref{fig:Fig1}(a) the schematic diagram of exciton states in two QDs coupled to a single-mode cavity and to a common phonon bath are shown. The QDs are pumped incoherently using external fields. We investigate the steady-state populations in collective states of QDs and cavity photon statistics. We also derive laser rate equation using the quantum theory of lasers developed by Scully and Lamb\cite{Scully1967,sargent1974}.
The Hamiltonian for the system in rotating frame is given by
\begin{equation}
    \begin{split}
     H =& \hbar \Delta_1 \sigma_1^+ \sigma_1^- + \hbar \Delta_2\sigma_2^+\sigma_2^- + \hbar(g_1\sigma_1^+ a+g_2\sigma_2^+ a+H.C)\\& + H_{ph},
    \end{split}
 \label{eqn:Ham}
\end{equation}
 where, the detuning $\Delta_i = \omega_i - \omega_c$, with $\omega_i$ and $\omega_c$ are the transition frequency between ground state $\ket{g_i}$ to excitonic state $\ket{e_i}$ for $i-{th}$ QD and cavity mode frequency, respectively. The lowering and raising operators for QDs are given by $\sigma_i^+ = \ket{e_i}\bra{g_i}$, $\sigma_i^- = \ket{g_i}\bra{e_i}$ and $g_i$ is the exciton-cavity mode coupling constant, $a$ is the cavity field operator. The last term in the Hamiltonian (\ref{eqn:Ham}) represents the exciton and longitudinal acoustic phonon interaction , $H_{ph} = \hbar\Sigma_k \omega_k b_k^\dagger b_k + \hbar\Sigma_i \lambda_k^i \ket{e_i}\bra{e_i}(b_k + b_k^\dagger)$. Here, $b_k$ is the field operator of phonon mode of frequency $\omega_k$ and $\lambda_k^i$ is the coupling strength of exciton $\ket{e_i}$ to the phonon mode. We derive the polaron transformed master equation to include exciton-phonon interaction in QD-cavity dynamics. Hence, we perform polaron transformation for the Hamiltonian(\ref{eqn:Ham}) using $H'=e^S H e^{-S}$, with $S=\Sigma_i \sigma_i^+\sigma_i^-\Sigma_k \frac{\lambda_k^{i}}{\omega_k}(b_k^\dagger-b_k)$. The transformed Hamiltonian can be written as $H' = H_s+H_b+H_{sb}$, where $H_s$ is  QD-cavity system Hamiltonian, $H_b$ is phonon-bath Hamiltonian and $H_{sb}$ is system-bath interaction Hamiltonian. 
\begin{equation}
H_s = \hbar\Delta_1\sigma_1^+\sigma_1^- + \hbar\Delta_2\sigma_2^+\sigma_2^- + \langle B \rangle X_g 
\label{eqn:Ham2}
\end{equation}
\begin{equation}
H_b = \hbar\Sigma_k\omega_k b_k^\dagger b_k
\end{equation}
\begin{equation}
H_{sb} = \zeta_g X_g + \zeta_u X_u 
\end{equation}
The polaron shifts, $\Sigma_k \frac{(\lambda_k^i)^2}{\omega_k}$ are absorbed in $\Delta_1, \Delta_2$. The phonon displacement operators are given by, $B_{\pm} = \exp[\pm \Sigma_k \frac{\lambda_k^{i}}{\omega_k}(b_k - b_k^\dagger]$, with $\langle B_{\pm} \rangle = \langle B \rangle$. For simplification, we have considered equal coupling strengths $\lambda_k^{1} = \lambda_k^{2}$. The system operators are, $X_g = \hbar(g_1\sigma_1^+a + g_2\sigma_2^+a)+H.C.$, $X_u = i\hbar(g_1\sigma_1^+a+g_2\sigma_2^+a)+H.C.$ and bath fluctuation operators are given by $\zeta_g = \frac{1}{2}(B_++B_- -2\langle B \rangle)$, $\zeta_u = \frac{1}{2i}(B_+ - B_-)$. Using the polaron transformed Hamiltonian, $H'$ and Born-Markov approximation, we derive the master equation for the QDs-cavity system \cite{roy2011}. The master equation for the density matrix of QDs-cavity system is given by,
\begin{equation}
    \begin{split}
   \dot{\rho_s} = &-\frac{i}{\hbar}[H_s,\rho_s]-L_{ph}\rho_s-\frac{\kappa}{2}L[a]\rho_s-\Sigma_{i=1,2}(\frac{\gamma_i}{2}L[\sigma_i^-]\\&+\frac{\gamma_i'}{2}L[\sigma_i^+\sigma_i^-]+\frac{\eta_i}{2}L[\sigma_i^+])\rho_s,
\end{split}
\label{eqn:incohME}
\end{equation}
where $L[\hat{O}]\rho = \hat{O^\dagger}\hat{O}\rho - 2\hat{O}\rho\hat{O^\dagger}+\rho\hat{O^\dagger}\hat{O}$ is the Lindblad super-operator. The second term in the master equation $L_{ph}\rho_s$ is written as 
\begin{equation}
\begin{split}
    L_{ph}\rho_s = &\frac{1}{\hbar^2}\int_{0}^{\infty}d\tau \Sigma_{j=g,u}G_j(\tau)\\&\times[X_j(t),X_j(t,\tau)\rho_s(t)]+H.C.
\end{split}
\label{eqn:Lph}
\end{equation}
where $X_j(t,\tau)=e^{-iH_s\tau/\hbar}X_j(t)e^{iH_s\tau/\hbar}$, and polaron Green's functions, $G_j(\tau)=\langle\zeta_j(t)\zeta_j(t,\tau)\rangle_{bath}$, $G_g(\tau)=\langle B \rangle^2{\cosh(\phi(\tau)-1)}$, $G_u(\tau)=\langle B \rangle^2\sinh(\phi(\tau))$. The phonon correlation function is given by,
\begin{equation}
    \phi(\tau)=\int_{0}^{\infty}d\omega\frac{J(\omega)}{\omega^2}[\coth(\frac{\hbar\omega}{2k_BT})\cos(\omega\tau)-i\sin(\omega\tau)],
\end{equation}
\par
where $k_B$, $T$ are the Boltzmann constant and temperature of the phonon bath, respectively. The spectral density function of phonon bath is given by $J(\omega)=\Sigma_k(\lambda_k^{e_i})^2\delta(\omega-\omega_k)=\alpha_p\omega^3\exp[-\frac{\omega^2}{2\omega_b^2}]$, takes the latter form in continuum limit. The electron-phonon coupling strength $\alpha_p$, depends on the deformation potential and the cut-off frequency $\omega_b$ depends on the speed of sound and phonon wave-function profile. Here we considered $\alpha_p=1.42 \times 10^{-3}g_1^2$, $\omega_b=10g_1$ which provide experimentally compatible values of $\langle B \rangle$=1.0, 0.9, 0.84 and 0.73 for $T$= 0K, 5K, 10K and 20K, respectively. We also include Lindblad terms corresponding to cavity damping with decay rate $\kappa$, spontaneous exciton decay with rate $\gamma_i$, pure dephasing with rate $\gamma_i'$ and incoherent pumping with rate $\eta_i$. The master equation (\ref{eqn:incohME}) is then numerically integrated to obtain the steady-state populations (SSP) and cavity photon statistics.

\begin{figure}
    \centering
    \includegraphics[width=\columnwidth]{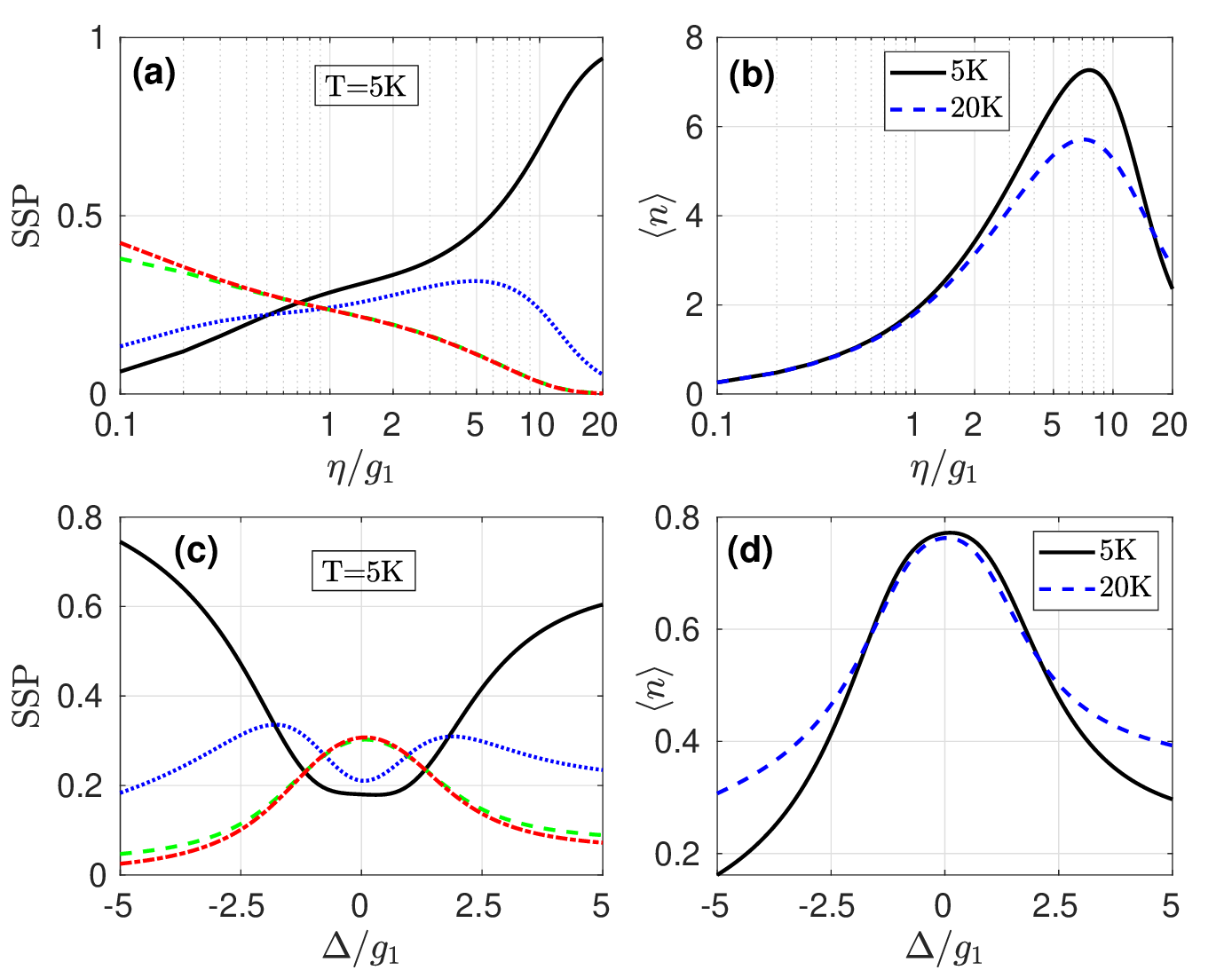}
    \caption{The steady-state populations in collective QD states, $\ket{e_1,e_2}$ (solid black), $\ket{+}=(\ket{e_1,g_2}+\ket{g_1,e_2})/\sqrt{2}$ (dotted blue), $\ket{-}=(\ket{e_1,g_2}-\ket{g_1,e_2})/\sqrt{2}$ (dashed green) and $\ket{g_1,g_2}$ (dash-dotted red) in (a) $\&$ (c). The cavity mean photon number, $\langle n\rangle$ for T=5K (solid black), T=20K (dashed blue) in (b) $\&$ (d). In (a) $\&$ (b) $\Delta_1 = \Delta_2=\Delta$, incoherent pumping rate, $\eta_1=\eta_2=0.35g_1$.   In (c) $\&$ (d) incoherent pumping rate $\eta_1=\eta_2=\eta$ and detuning, $\Delta_1=\Delta_2=\Delta=0.0$. Others parameters are cavity decay rate $\kappa=0.5g_1$, spontaneous emission rate $\gamma_1=\gamma_2=0.01g_1$, pure dephasing rate $\gamma_1'=\gamma_2'=0.01g_1$.}
    \label{fig:Fig2}
\end{figure}

\begin{figure}
    \centering
    \includegraphics[width=\columnwidth]{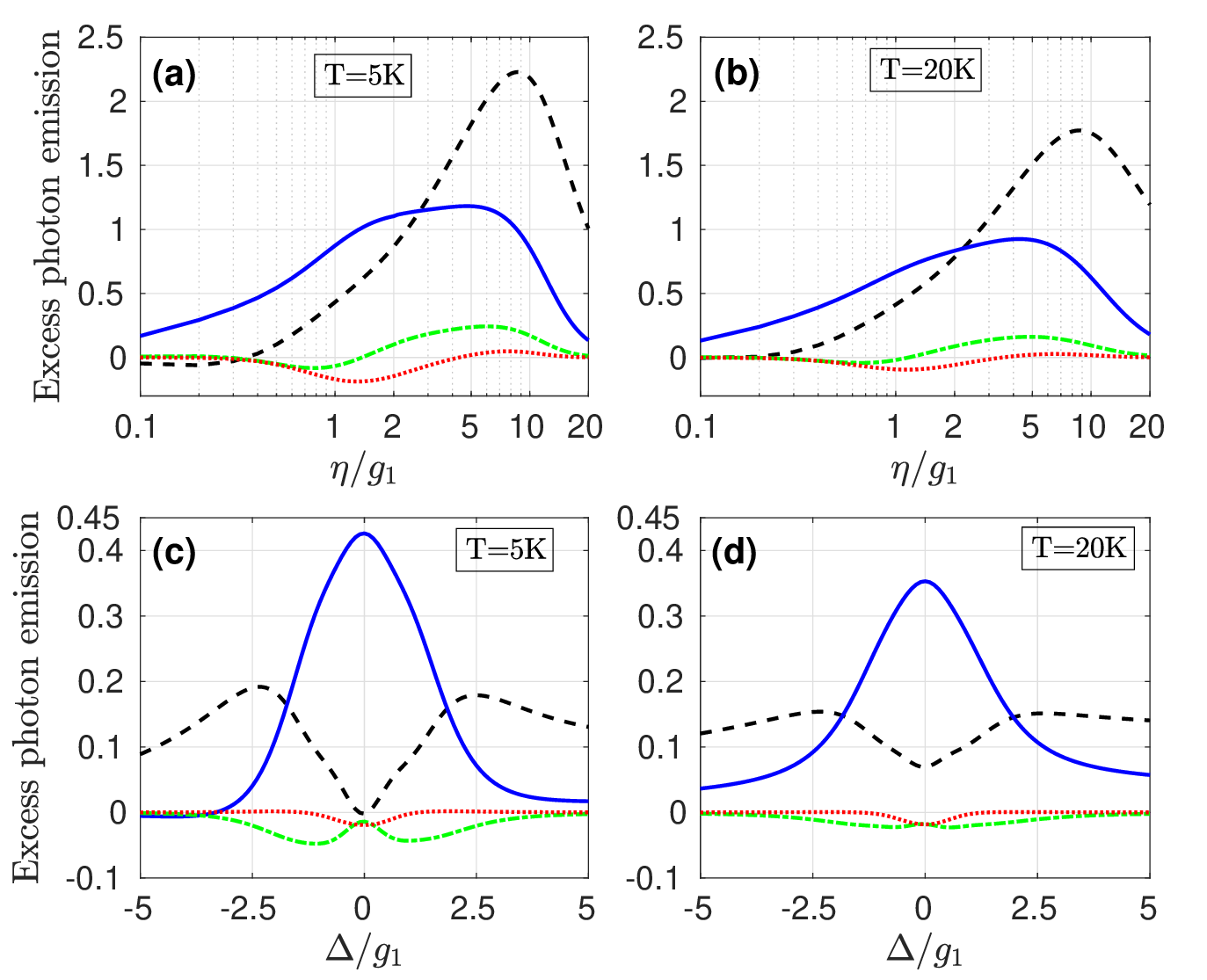}
    \caption{Excess photon emission at T=5K, (a) $\&$ (c), and at T=20K, (b) $\&$ (d). Single photon excess emission(SPEE, dashed black), Two photon excess emission(TPEE, solid blue), Three photon excess emission (ThPEE, dotted red), Four photon excess emission (FPEE, dash-dotted green). The parameters are same as in Fig.\ref{fig:Fig2}.}
    \label{fig:Fig3}
\end{figure}

 The steady-state populations(SSP) in two-QD states and average photons in cavity mode are shown in Fig.\ref{fig:Fig2}.  In Fig.\ref{fig:Fig2}(a) and (b) the steady state populations and mean cavity photon number with respect to incoherent pumping rate, $\eta_1=\eta_2=\eta$, while the QDs are resonantly coupled to cavity mode $\Delta_1=\Delta_2=0$ are presented. In Fig.\ref{fig:Fig2} (c) and (d) the results  with respect to detuning, $\Delta_1=\Delta_2=\Delta$ for a fixed value of incoherent pumping rate $\eta_1=\eta_2=\eta$ are shown. We notice that two proximate QDs can be tuned to resonance using electric field\cite{Koong2022}. 
 Fig. \ref{fig:Fig2}(a) $\&$ (b) show the results corresponding to variation in the incoherent pumping rate, considering both QDs are driven equally, $\eta_1=\eta_2=\eta$ and are resonantly coupled to cavity mode, $\Delta_1=\Delta_2=\Delta=0$. From Fig. \ref{fig:Fig2}(a), we find that the steady state population in state $\ket{e_1,e_2}$ start dominating for very small pump rate, $\eta<g_1$ and increases on increasing pump rate. For $g_1=g_2$, the state $\ket{+}=(\ket{e_1,g_2}+\ket{g_1,e_2})/\sqrt{2}$ is coupled with cavity mode and the state $\ket{-}=(\ket{e_1,g_2}-\ket{g_1,e_2})/\sqrt{2}$ remains uncoupled with cavity mode. The population in $\ket{+}$ increases monotonically on increasing pump rate. The state $\ket{-}$ is populated due to incoherent pumping and spontaneous decay of $\ket{e_1,e_2}$.
 Further, the steady state populations in collective QD states, $\ket{g_1,g_2}$ and $\ket{-}$ remain equal and decrease monotonically on increasing pump strength. The collective QD states $\ket{e_1,e_2}$, $\ket{+}$, and $\ket{g_1,g_2}$ get dressed with the cavity field as shown in Fig.\ref{fig:Fig1}(c) for $\Delta=0$. It should also be noted that at higher pump rate, the dressed states with higher number of cavity photons, get more populated. Various possible transitions between these dressed states can lead to the single and multi-photon emission into cavity mode as shown in Fig.\ref{fig:Fig1}(b). 
 The average photon number in cavity mode, $\langle n\rangle$, increases and attains a peak value on increasing pump rate, as shown in Fig. \ref{fig:Fig2}(b). On further increasing the incoherent pump rate, the mean photon number decreases due to the destruction of coherence in the system and leading to self-quenching effect \cite{mu1992}. The self-quenching also leads to rapid increase and decrease in the populations of $\ket{e_1,e_2}$ and $\ket{+}$ states, respectively. For higher temperature, T=20K, steady-state populations and cavity photon statistics follow a similar fashion. In Fig. \ref{fig:Fig2}(b), the peak in $\langle n\rangle$ (dashed blue), is smaller for T=20K than for T=5K (solid black) due to renormalization of cavity QD coupling by a factor $\langle B\rangle$.
In Fig. \ref{fig:Fig2}(c) $\&$ (d) we plot steady state populations and average cavity photon with the QDs detunings $\Delta_1=\Delta_2=\Delta$, for typical value of pump rate, $\eta_1=\eta_2=\eta=0.35g_1$.  We choose pump rate corresponding to the single-photon excessive emission becomes zero (c.f. Fig. \ref{fig:Fig3}(a)). 
In Fig. \ref{fig:Fig2}(c), we find that when the QDs are tuned to resonance with the cavity mode, $\Delta_1=\Delta_2=\Delta=0$, the populations in excited states $\ket{e_1,e_2}, \ket{+}$ deplete and the population in ground state $\ket{g_1,g_2}$ increases. The average photon number in cavity mode, $\langle n\rangle$ also has peak when QDs are resonant with cavity mode, shown in Fig. 2(d). The asymmetry in the mean photon number curve is due to phonon-induced cavity mode feeding, that is more pronounced for positive detunings. Clearly, it suggests that QDs undergo transitions, $\ket{e_1,e_2}\rightarrow\ket{+}\rightarrow\ket{g_1,g_2}$ accompanied with two-photon emission into cavity mode. The anti-symmetric state, $\ket{-}$ is more populated than $\ket{+}$ when QDs are at resonance, $\Delta=0$ for low pumping rate. Comparing the results for the T=20K case with T=5K, the peak value of average photon number $\langle n\rangle$ is slightly reduced when the QDs are resonant, but the values are higher at higher temperature for off-resonant QDs due to phonon assisted transitions. 

In order to quantify photon emission into the cavity mode from single and multi-photon processes, we derive laser rate equation using simplified master equation(SME). The SME is approximated Lindblad form of polaron transformed master equation (\ref{eqn:incohME}), which provides a clear picture of the various exciton-phonon scattering processes involved in the system. For deriving SME, we expand $L_{ph}\rho_s$, after making approximation $H_s\approx\hbar\Delta_1 \sigma_1^+\sigma_1^-+\hbar\Delta_2\sigma_2^+\sigma_2^-$ in $X_j(t,\tau)=e^{-iH_s\tau/\hbar}X_j(t)e^{iH_s\tau/\hbar}$. We arrange terms in the Lindblad form which are proportional to $g_1^2, g_2^2$ and $g_1g_2$ in $L_{ph}\rho_s$\cite{verma2018}.
The simplified master equation is given by,
\begin{equation}
 \begin{split}
  \dot{\rho_s}=& -\frac{i}{\hbar}[H_{eff},\rho_s]-\frac{\kappa}{2}L[a]\rho_s
     -\Sigma_{i=1,2}(\Big[\frac{\eta_i}{2}L[\sigma_i^+])
    \\&+\frac{\gamma_i}{2}L[\sigma_i^-]
    +\frac{\gamma_i'}{2}L[\sigma_i^+\sigma_i^-]
    +(\frac{\Gamma_i^-}{2}L[\sigma_i^+a]+H.C)\Big]\rho_s
    \\&-\Big[\frac{\Gamma_{12}^{--}}{2}L[\sigma_2^+a,\sigma_1^+a]\rho_s
    +\frac{\Gamma_{12}^{++}}{2}L[a^\dagger\sigma_2^-,a^\dagger\sigma_1^-]\rho_s
    \\&+\frac{\Gamma_{12}^{-+}}{2}L[a^\dagger\sigma_2^-,\sigma_1^+a]\rho_s
    +\frac{\Gamma_{12}^{+-}}{2}L[\sigma_2^+a,a^\dagger\sigma_1^-]\rho_s
    \\&+\Omega_{11}^{--}\sigma_1^+ a\rho_s\sigma_1^+a +\Omega_{11}^{++}\sigma_1^-a^\dagger\rho_s\sigma_1^-a^\dagger
    +1 \leftrightarrow 2 \Big]
    \label{eqn:incohSME}
    \end{split}
 \end{equation}
Here, $L[\hat{O_1},\hat{O_2}]=\hat{O_2}\hat{O_1}\rho_s-2\hat{O_1}\rho_s\hat{O_2}+\rho_s\hat{O_1}\hat{O_2}$. The coherent evolution of the system density matrix, $\rho_s$ is given by the effective Hamiltonian, $H_{eff}$
\begin{equation}
\begin{split}
    H_{eff} = &H_s+\hbar\Sigma_{i=1,2}(\delta_i^-a^\dagger\sigma_i^-\sigma_i^+a+\delta_i^+\sigma_i^+a a^\dagger\sigma_i^-)
    \\&-(i\hbar\Omega_{2ph}\sigma_1^+\sigma_2^+a^2+H.C.)-(i\hbar\Omega_+\sigma_1^+a a^\dagger\sigma_2^- \\&+i\hbar\Omega_-a^\dagger\sigma_1^-\sigma_2^+a+H.C.)
\end{split}
\end{equation}
where $\delta_i^\pm$, $\Omega_{2ph}$, $\Omega_\pm$ correspond to the stark shifts, two-photon interaction and exciton exchange between QDs, respectively. Further, $\Gamma_i^\pm$, $\Gamma_{ij}^{++}$, $\Gamma_{ij}^{--}$, $\Gamma_{ij}^{+-}$, $\Gamma_{ij}^{-+}$ are phonon-induced scattering rates of the incoherent processes such as cavity mode feeding, two-photon emission, two-photon absorption, exciton exchange, respectively. We also include terms including $\Omega_{ii}^{\pm\pm}$ that do not have Lindblad form. The terms discussed above are defined as
\begin{equation}
    \delta_i^\pm=g_i^2Im\left[\int_0^\infty d\tau G_+e^{\pm i\Delta_i\tau}\right]
\end{equation}
\begin{equation}
    \Omega_{2ph} = \frac{g_1g_2}{2}\int_0^\infty d\tau(G_- - G_-^*)(e^{-i\Delta_1\tau}+e^{-i\Delta_2\tau})
\end{equation}
\begin{equation}
    \Omega_\pm = \frac{g_1g_2}{2}\int_0^\infty d\tau(G_+e^{\pm i\Delta_2\tau}-G_+^*e^{\mp i\Delta_1\tau})
\end{equation}
\begin{equation}
    \Gamma_i^\pm = g_i^2\int_0^\infty d\tau(G_+e^{\pm i\Delta_i\tau}+G_+^*e^{\mp i\Delta_i\tau})
\end{equation}
\begin{equation}
    \Gamma_{ij}^{\pm\pm} = g_ig_j\int_0^\infty d\tau(G_-e^{\pm i\Delta_j\tau}+G_-^*e^{\pm i\Delta_i\tau})
\end{equation}
\begin{equation}
    \Gamma_{ij}^{\pm \mp} = g_ig_j\int_0^\infty d\tau(G_+e^{\mp i\Delta_j\tau}+G_+^*e^{\pm i\Delta_i\tau})
\end{equation}
\begin{equation}
    \Omega_{ii}^{\pm\pm} = g_i^2 \int_0^\infty d\tau(G_- + G_-^*)e^{\pm i\Delta_i\tau}
\end{equation}
 We compare steady state populations and $\langle n\rangle$ calculated using master equation, (5) and SME (\ref{eqn:incohSME}), we find that SME is valid entirely in the range of detunings and pumping rate considered here (c.f. Fig.\ref{fig:Fig7}(a), (b)).

In order to obtain the quantum laser rate equation for the cavity field, we write the rate equations for both diagonal and off-diagonal QDs-cavity density matrix elements using SME (\ref{eqn:incohSME}). Following quantum theory of lasers developed by Scully and Lamb, we express off-diagonal elements in terms of diagonal elements under steady state condition. After tracing over collective QD states, we obtain rate equation for probability of having `n' photons in the cavity $P_n=P_n^{ee}+P_n^{eg}+P_n^{ge}+P_n^{gg}$, where $P_n^{ab}=\langle a_1,b_2,n|\rho_s|a_1,b_2,n\rangle$; with $a,b=e,g$. The laser rate equation is given by
\begin{widetext}
    \begin{equation}
    \begin{split}
    \dot{P_n}=&-[\alpha_n^{ee}P_n^{ee}+\alpha_n^{eg}P_n^{eg}+\alpha_n^{ge}P_n^{ge}+\alpha_n^{gg}P_n^{gg}]+\sum_{k=1}^m (\Gamma_{n+k}^{ee(k)}P_{n+k}^{ee}+\Gamma_{n+k}^{eg(k)}P_{n+k}^{eg}+\Gamma_{n+k}^{ge(k)}P_{n+k}^{ge}+\Gamma_{n+k}^{gg(k)}P_{n+k}^{gg})\\&+\sum_{k=1}^m (G_{n-k}^{ee(k)}P_{n-k}^{ee}+G_{n-k}^{eg(k)}P_{n-k}^{eg}+G_{n-k}^{ge(k)}P_{n-k}^{ge}+G_{n-k}^{gg(k)}P_{n-k}^{gg})-\kappa nP_n+\kappa (n+1)P_{n+1}
    \end{split}
    \label{eqn:cavityRE}
\end{equation}
\end{widetext}
The second term in Eq.\eqref{eqn:cavityRE} corresponds to k-photon absorption, and the third term corresponds to k-photon emission in the cavity mode. The coefficients $\alpha^{ab}_n$, $\Gamma^{ab(k)}_n$, and $G^{ab(k)}_n$ are calculated numerically. We consider up to four-photon processes in laser rate equation, therefore truncate the summation in Eq.\eqref{eqn:cavityRE} for $m=4$. The terms for $m>4$ remain negligible. Here we do not use mean field approximation to separate QD-cavity correlations. Further, we find $\sum_{a}\sum_{b}\alpha_n^{ab}P_n^{ab}=\sum_{k=1}^m\sum_{a}\sum_{b}[\Gamma_n^{ab(k)}P_n^{ab}+G_n^{ab(k)}P_n^{ab}]$. The photon emission into cavity mode occurs via both stimulated emission and spontaneous emission similar to high $\beta$ laser\cite{Li2009}. Using the laser rate equation \eqref{eqn:cavityRE}, the steady state mean photon number in cavity mode is given by
\begin{widetext}
\begin{equation}
    \langle n\rangle = \frac{\sum_{n}\sum_{k} k\Big[(G_n^{ee(k)}-\Gamma_n^{ee(k)})P_n^{ee}+(G_n^{eg(k)}-\Gamma_n^{eg(k)})P_n^{eg}+(G_n^{ge(k)}-\Gamma_n^{ge(k)})P_n^{ge}+(G_n^{gg(k)}-\Gamma_n^{gg(k)})P_n^{gg})\Big]}{\kappa}.
    \label{eqn:meanphotonNumber}
\end{equation}
\end{widetext}

In Eq.\eqref{eqn:meanphotonNumber}, the terms corresponding to $k=1,2,3,4$ in average cavity photons are corresponding to single photon excessive emission, two-photon excessive emission, three-photon excessive emission, four-photon excessive emission into cavity mode. The positive value of $k$-th term implies there is net emission into cavity mode through $k$-photon process and the negative value implies absorption from cavity mode through $k$-photon process. We have also compared the average photon numbers in cavity mode, $\langle n\rangle$ obtained from \eqref{eqn:meanphotonNumber} considering up to four-photon processes and the values obtained from the SME \eqref{eqn:incohSME} in the steady-state; the values match very well (c.f. Fig.\ref{fig:Fig8}(a)).

We plot the excess photon emission into cavity mode via single and multi-photon processes in Fig.\ref{fig:Fig3}.
Considering QDs are resonantly coupled with cavity mode, in Fig.\ref{fig:Fig3} (a) $\&$ (b), the results of excess photon emission into cavity mode up to four-photon processes for T=5K and 20K, respectively, with increasing incoherent pumping rate $\eta$, are presented. The transitions, between dressed states of QDs with more than one photon (Fig.\ref{fig:Fig1}(c)), lead to multi-photon absorption and emission in the cavity mode.
For low pumping rate ($\eta\leq 2g_1$), two-photon excessive emission into the cavity mode dominates single-photon excessive emission. A small three-photon and four-photon excessive absorption and emission are also present. We find that when pump strength is varied, the single and multi-photon excess emissions change from negative to positive values at different pump values. Especially, for $\eta=0.35g_1$, excess single photon emission is zero and the cavity mode is populated only due to two-photon emission and the system behaves as a two-photon laser. For higher pump rate, single-photon excessive emission grows rapidly compared to multi-photon excess emission into cavity mode. This domination of single photon emission over others shows that emission from individual QDs dominates over cooperative emission at higher pump strength. On further increasing pumping rate, the excessive emission into cavity mode decreases as a result of self quenching which is also evident in mean photon number in cavity mode, in Fig. 2(b). For higher temperature at T=20K, in Fig. 3(b), phonon induced dephasing rises, leading to decrease in the emission into cavity mode. However, single-photon excessive emission increases for low pumping rates, $\eta<g_1$. We also find that the domination of single-photon excess emission over two-photon excess emission occurs at smaller pump rate.
Therefore, it is possible by appropriately choosing pump rate, the emission in the cavity mode occurs from cooperative two-photon processes predominantly.
We present the results by varying the QDs detuning, $\Delta$ with respect to cavity mode in Fig. 3(c) $\&$ (d), for pump rate when single-photon excess emission in cavity mode becomes zero. It is observed that when both QDs are coupled resonantly with cavity mode, two-photon excess emission dominates indicating dominating two-photon lasing in the system. However, when QDs are off-resonant single-photon excessive emission dominates. Further, a small three-photon and four-photon absorption also appear at low temperature when QDs are resonant to the cavity mode. For higher temperature at T=20K, at $\Delta=0$, the two-photon excessive emission decreases but single-photon excessive emission increases as discussed earlier. Increased exciton-phonon scattering at higher temperature diminishes the three-photon and four-photon absorption as well. We also notice broadening of excess photon emission curves similar to the curves for $\langle n\rangle$ in Fig. \ref{fig:Fig2}(c).

\section{Cooperative two-photon lasing in coherently pumped two-QD}

In this section we consider coherently pumped two-QD coupled with a single mode photonic microcavity. The Hamiltonian of the system in rotating frame with pump frequency is written as,
\begin{equation}
    \begin{split}
    \hat{H}=&\hbar\Delta_{cp}a^\dagger a+\hbar\Delta_{1p}\sigma_1^+\sigma_1^-+\hbar\Delta_{2p}\sigma_2^+\sigma_2^-\\&+\hbar(g_1\sigma_1^+a+g_2\sigma_2^+a+H.C)\\&+\hbar(\eta_1\sigma_1^++\eta_2\sigma_2^++H.C)+\hat{H}_{ph}
    \end{split}
\end{equation}
where $\eta_i$ is the coupling strength of coherent field with $i^{th}$ QD. The detuning of the cavity mode, and the detunings of $i^{th}$ QD with respect to the pump frequency are $\Delta_{cp}=\omega_c-\omega_p$, and $\Delta_{ip}=\omega_i-\omega_p$, respectively.
 As discussed in the previous section, for including exciton-phonon interaction non-perturbatively we construct the polaron transformed master equation (5), where the system Hamiltonian and the system operators are given by,
\begin{equation}
    H_s=\hbar\Delta_{cp}a^\dagger a+\hbar\Delta_{1p}\sigma_1^+\sigma_1^-+\hbar\Delta_{2p}\sigma_2^+\sigma_2^-+\langle B \rangle X_g,
\label{hs_coh}
\end{equation}
\begin{equation}
X_g=\hbar(g_1\sigma_1^+a+g_2\sigma_2^+a+\eta_1\sigma_1^+\eta_2\sigma_2^+)+H.C,
\label{xg_coh}
\end{equation}
\begin{equation}
X_u=i\hbar(g_1\sigma_1^+a+g_2\sigma_2^+a+\eta_1\sigma_1^+\eta_2\sigma_2^+)+H.C.
\label{xu_coh}
\end{equation}
We use master equation (5), with system operators (\ref{hs_coh}), (\ref{xg_coh}), and (\ref{xu_coh}) to calculate the steady state population in QDs and the average photons in cavity mode.
\begin{figure}
    \centering
    \includegraphics[width=\columnwidth]{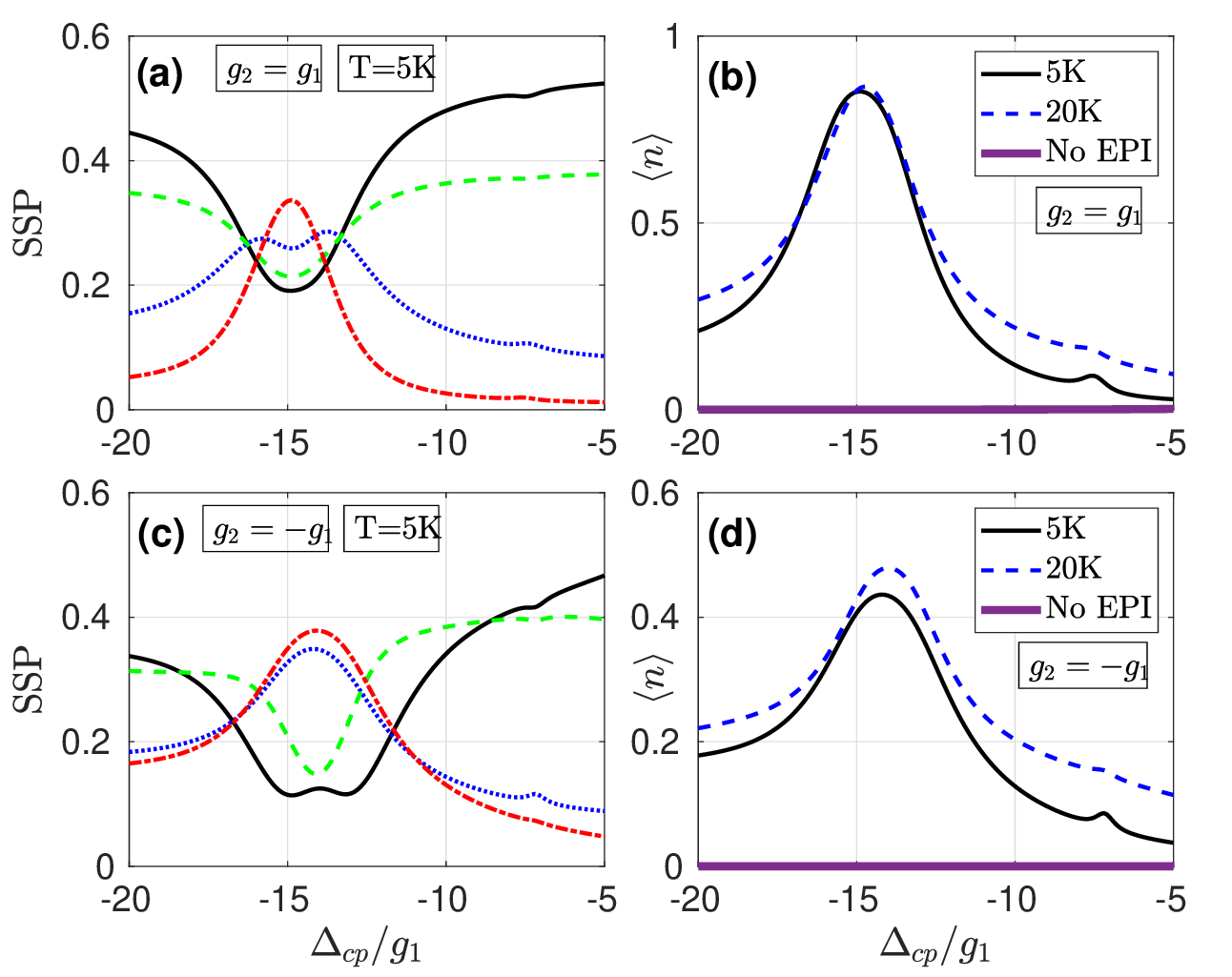}
    \caption{The steady state populations in QD states and cavity photon number for coherently driven system, (a), (b) for the case of $g_2=g_1$ and (c), (d) for $g_2=-g_1$. (a) $\&$ (c) show steady state populations for T=5K, (b) $\&$ (d) show mean cavity photon number, $\langle n\rangle$ for the cases T=5K, T=20K, No EPI(thick solid violet) by varying the cavity detuning $\Delta_{cp}$, considering the other parameters $\kappa=0.5g_1, \Delta_{1p}=\Delta_{2p}=-13.5 g_1$, $\eta_1=\eta_2=\eta=3.0 g_1$ in (a),(b), $\eta_1=\eta_2=\eta=1.9g_1$ in (c),(d) $\gamma_1=\gamma_2=0.01g_1, \gamma'_1=\gamma'_2=0.01g_1 $. Color scheme is same as in Fig.\ref{fig:Fig2} (a) $\&$ (b).}
    \label{fig:Fig4}
\end{figure}
The results are shown in Fig.\ref{fig:Fig4}(a),(b) when both QDs are symmetrically ($g_2= g_1$) coupled and in Fig.\ref{fig:Fig4}(c),(d) when both QDs are asymmetrically ($g_2=-g_1$) coupled to the cavity mode. Earlier, the population inversion in QDs has been demonstrated for strong blue detuned coherent pump due to phonon-assisted transitions\cite{Quilter2015}. Therefore, we consider off-resonant coherent pump $\Delta_{1p}=\Delta_{2p}=\Delta=-13.5g_1$. We find that resonant transitions in cavity mode occur when $\Delta_{cp}=-\Omega'$ where  $\Omega'=\sqrt{\Delta_{ip}^2+4\eta^2}$ is frequency difference between QD states dressed with coherent pump. We have chosen optimal pumping rates $\eta=3.0g_1$, $\eta=1.9g_1$ for symmetric, anti-symmetric cases respectively for the system to act as two photon laser as shown latter in this section. Therefore, $\Omega'=14.75g_1$ in (a) $\&$ (b) for $\eta=3.0g_1$ and $\Omega'=14.0g_1$ in (c) $\&$ (d) for $\eta=1.9g_1$.
In Fig.\ref{fig:Fig4}(a) $\&$ (c) we see a dip in the steady state population of $\ket{e_1,e_2}$ and a peak in the population of $\ket{g_1,g_2}$ at $\Delta_{cp}=\Omega'$ indicating resonant transitions in cavity mode.
Also, there is a dip in the populations of state $\ket{-}$ and the population of state $\ket{+}$ is greater than the population of state $\ket{-}$. The changes in the populations of states $\ket{+}$ and $\ket{-}$ occur due to single-photon coherent and incoherent transitions. Further, for $g_2=g_1$ the state $\ket{+}$ is coupled with cavity mode and the state $\ket{-}$ remains uncoupled. Therefore, change in population of $\ket{-}$ occurs due to incoherent single-photon transitions only, leading to broader dip around lasing transitions (Fig.\ref{fig:Fig4}(a)). Similarly, for $g_2=-g_1$ the state $\ket{+}$, is coherently coupled with pump but remains uncoupled with the cavity mode leading to a broader peak around the resonance (Fig.\ref{fig:Fig4}(c)). We also observe a tiny dip in the steady state population of $\ket{e_1,e_2}$ and a tiny peak in the population of $\ket{+}$ for $\Delta_{cp}=-\Omega'/2$, where cavity is resonant with two-photon transition between pump dressed states of individual QDs. Such transitions have been utilized to generate two-photon dressed state laser\cite{Gauthier1992}. Corresponding to the cavity detuning $\Delta_{cp}=-\Omega'$, in Fig.\ref{fig:Fig4}(b) and (d), there is maxima in the average photon number in the cavity mode. There is also a small peak in average number of cavity photons corresponding to $\Delta_{cp}=-\Omega'/2$. Further there is no major difference in cavity photons whether QDs are symmetrically ($g_1=g_2$) or anti-symmetrically ($g_1=-g_2$) coupled with the cavity mode. With increase in temperature, at T=20K, there is more emission in cavity mode due to off-resonant phonon assisted cavity mode feeding \cite{Hohenester2010} and the peaks are broadened. We also notice that without including exciton-phonon interaction the emission in cavity mode is negligible, because without phonon interactions off-resonant pump could not generate significant population in exciton state. The steady state mean photon number in cavity mode can be analysed by considering the single and multi-photon excessive emission in cavity mode using laser rate equation.

Following the similar approach as discussed in previous section, we derive laser rate equation \eqref{eqn:cavityRE}. We use simplified master equation for deriving laser rate equation. In order to construct the simplified master equation, we approximate system Hamiltonian, $H_s\approx\hbar\Delta_{cp}a^\dagger a+\hbar\Delta_{1p}\sigma_1^+\sigma_1^-+\hbar\Delta_{2p}\sigma_2^+\sigma_2^-$ in $L_{ph}\rho_s$ and arranging terms proportional to $g_1^2$, $g_2^2$, $g_1g_2$, $\eta_1^2$,$\eta_2^2$, and $\eta_1\eta_2$ is given by
\begin{widetext}
\begin{equation}
    \begin{split}
    \dot{\rho_s}=&-\frac{i}{\hbar}[H_{eff},\rho_s]-\frac{\kappa}{2}L[a]\rho_s-\Sigma_{i=1,2}(\frac{\gamma_i}{2}L[\sigma_i^-]+\frac{\gamma_i'}{2}L[\sigma_i^+\sigma_i^-])\rho_s+\Sigma_{i=1,2}(\frac{\Gamma_i^-}{2}L[\sigma_i^+a]+\frac{\Gamma_i^+}{2}L[\sigma_i^-a]+\frac{\Gamma_p^{\sigma_i^+}}{2}L(\sigma_i^+)+\\&\frac{\Gamma_p^{\sigma_i^-}}{2}L(\sigma_i^-))\rho_s
    -\Big[(\frac{\Gamma_{12}^{--}}{2}L[\sigma_2^+a,\sigma_1^+a]+\frac{\Gamma_{12}^{++}}{2}L[\sigma_2^-a^\dagger,\sigma_1^-a^\dagger]+\frac{\Gamma_{12}^{-+}}{2}L[a^\dagger\sigma_2^-,\sigma_1^+a]+
    +\frac{\Gamma_{12}^{+-}}{2}L[\sigma_2^+a,a^\dagger\sigma_1^-])\rho_s
    \\&+\Omega_{11}^{++}\sigma_1^+ a\rho_s\sigma_1^+ a+\Omega_{11}^{--}\sigma_1^-a^\dagger\rho_s\sigma_1^-a^\dagger
    +1 \leftrightarrow 2 \Big]-\Big[\sum\limits_{k,l=\pm}\frac{\Gamma_p^{\sigma_2^k\sigma_1^l}}{2}L[\sigma_2^k,\sigma_1^l]
    +\Omega_p^{\sigma_1^+\sigma_1^+}\sigma_1^+\rho_s\sigma_1^+ + H.C.
    +1 \leftrightarrow 2 \Big]
    \end{split}
    \label{eqn:cohSME}
\end{equation}
\end{widetext}
The coherent evolution of the system is given by $H_{eff}$, which includes phonon-mediated stark shifts, $\delta_i^\pm$, $\delta_p^{\sigma_i^\pm}$, two-photon processes corresponding to $\Omega_{2ph}$, $\Omega_p^{++}$, phonon-assisted exciton exchange processes corresponding to $\Omega_\pm$, $\Omega_p^\pm$.
The effective Hamiltonian of the system is given by
\begin{equation}
\begin{split}
    H_{eff} = &H_s+\hbar\Sigma_{i=1,2}(\delta_i^-a^\dagger\sigma_i^-\sigma_i^+a+\delta_i^+\sigma_i^+a a^\dagger\sigma_i^-)
    \\&-(i\hbar\Omega_{2ph}\sigma_1^+\sigma_2^+a^2+H.C.)-(i\hbar\Omega_+\sigma_1^+a a^\dagger \\&+i\hbar\Omega_-a^\dagger\sigma_1^-\sigma_2^+a+H.C.)+\hbar\Sigma_{i=1,2}\delta_p^{\sigma_i^+}\sigma_i^+\sigma_i^-\\&+i\hbar\Sigma_{i=1,2}\delta_p^{\sigma_i^-}\sigma_i^-\sigma_i^+ +(i\hbar\Omega_p^{++}\sigma_1^+\sigma_2^++H.C.)\\&-(i\hbar\Omega_p^+\sigma_1^+\sigma_2^-+i\hbar\Omega_p^-\sigma_1^-\sigma_2^++H.C.),
\end{split}
\end{equation}
with
\begin{equation}
    \delta_p^{\sigma_i^\pm}=\eta_i^2Im[\int_0^\infty d\tau G_+e^{\pm i\Delta_{ip}\tau}],
\end{equation}
\begin{equation}
    \Omega_p^{++} = \frac{\eta_1\eta_2}{2}\int_0^\infty d\tau(G_- - G_-^*)(e^{-i\Delta_{1p}\tau}+e^{-i\Delta_{2p}\tau}),
\end{equation}
\begin{equation}
    \Omega_p^\pm = \frac{\eta_1\eta_2}{2}\int_0^\infty d\tau(G_+e^{\pm i\Delta_{2p}\tau}-G_+^*e^{\mp i\Delta_{1p}\tau}).
\end{equation}
The additional terms  in the above simplified master equation (\ref{eqn:cohSME}), apart from the ones present in the incoherent case (\ref{eqn:incohSME}), includes phonon-assisted incoherent excitation corresponding to $\Gamma_p^{\sigma_i^+}$, enhanced exciton decay process corresponding to $\Gamma_p^{\sigma_i^-}$, double exciton creation and annihilation terms proportional to $\Gamma_p^{\sigma_2^+\sigma_1^+}$, $\Gamma_p^{\sigma_2^-\sigma_1^-}$ and exciton transfer proportional to $\Gamma_p^{\sigma_2^-\sigma_1^+}$, $\Gamma_p^{\sigma_2^+\sigma_1^-}$. The expressions of rates corresponding to these processes are 
\begin{equation}
    \Gamma_p^{\sigma_i^\pm} = \eta_i^2 \int_0^\infty d\tau(G_+e^{\pm i\Delta_{ip}\tau}+G_+^*e^{\mp i\Delta_{ip}\tau}),
\end{equation}
\begin{equation}
    \Gamma_p^{\sigma_i^\pm \sigma_j^\pm} = \eta_i\eta_j\int_0^\infty d\tau(G_-e^{\pm i\Delta_{ip}\tau} + G_-^*e^{\pm i\Delta_{jp}\tau}),
\end{equation}
\begin{equation}
    \Gamma_p^{\sigma_i^\mp \sigma_j^\pm} = \eta_i\eta_j\int_0^\infty d\tau(G_+e^{\pm i\Delta_{ip}\tau}+ G_+^* e^{\mp i\Delta_{jp}\tau}),
\end{equation}
\begin{equation}
    \Omega_p^{\sigma_i^\pm \sigma_i^\pm} = \eta_i^2 \int_0^\infty d\tau(G_- + G_-^*)e^{\mp i\Delta_{ip}\tau}.
\end{equation}
We have also retained some of the terms proportional to $\Omega_p^{\sigma_i^\pm \sigma_i^\pm}$, which do not have Lindblad form. However, these terms are necessary for better approximation. The comparison between polaron transformed master equation \eqref{eqn:incohME} and simplified master equation \eqref{eqn:cohSME} is relegated to the Appendix A.

\begin{figure}
    \centering
    \includegraphics[width=\columnwidth]{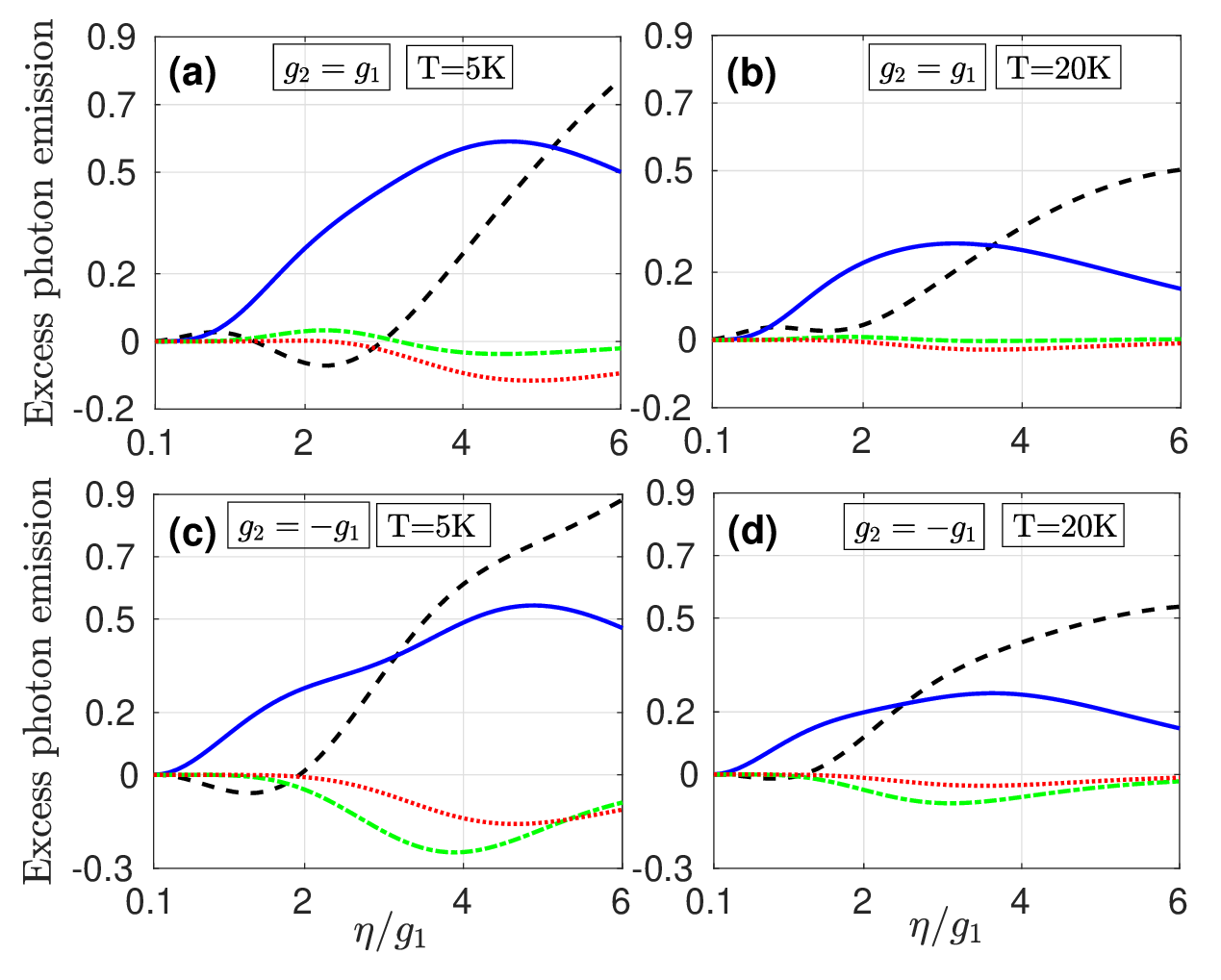}
    \caption{Single and multi excess photon emission for $\Delta_{1p}=\Delta_{2p}=-13.5 g_1, \Delta_{cp}=\Omega'$ with varying coherent pumping rate, $\eta_1=\eta_2=\eta$ and other parameters are same as in Fig.\ref{fig:Fig4} for the cases, (a) $g_2=g_1$, T=5K (b) $g_2=g_1$, T=20K (c) $g_2=-g_1$, T=5K (d) for $g_2=-g_1$, T=20K. Color scheme is same as in Fig. \ref{fig:Fig3}.}
    \label{fig:Fig5}
\end{figure}
\begin{figure}
    \centering
    \includegraphics[width=\columnwidth]{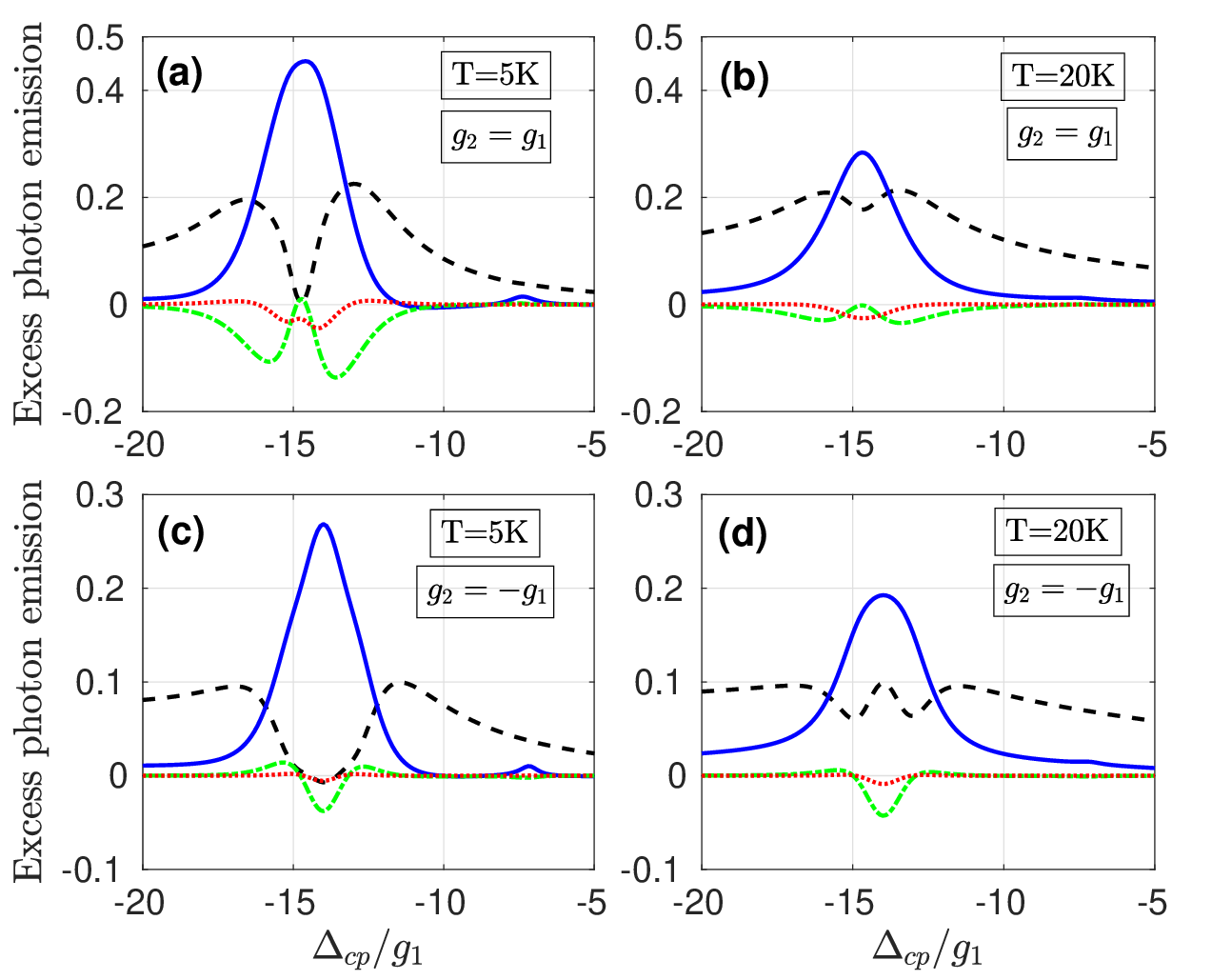}
    \caption{Excess photon emission for (a) $g_2=g_1$, T=5K (b) $g_2=g_1$, T=20K (c) $g_2=-g_1$, T=5K (d) for $g_2=-g_1$, T=20K rest all parameters are same as in Fig. \ref{fig:Fig4}(a). Color scheme is same as in Fig.\ref{fig:Fig3}.}
    \label{fig:Fig6}
\end{figure}

Following the laser rate equation \eqref{eqn:cavityRE}, we present single and multi-photon excessive emission into cavity mode in Fig.\ref{fig:Fig5} and Fig.\ref{fig:Fig6}.
 In Fig. \ref{fig:Fig5} we show the results by varying the coherent pumping rate, $\eta_1=\eta_2=\eta$. We consider far off-resonant blue detuned coherent pump and cavity detuning, $\Delta_{cp}=-\Omega'$, so that cavity becomes resonant with transition between pump dressed QD states. In Fig. \ref{fig:Fig5}(a) $\&$ (b), we consider the QDs are coupled symmetrically ($g_1=g_2$) and in Fig. \ref{fig:Fig5}(c) $\&$ (d) QDs are coupled anti-symmetrically ($g_2=-g_1$) to the cavity mode. The symmetric and anti-symmetric coupling of QDs with cavity mode leads to different behaviour as both coherent pump and cavity mode couple to $\ket{+}$ state in first case, whereas  coherent pump couples to $\ket{+}$ state but cavity couples to $\ket{-}$ state resulting in distinct interference effects between transitions \cite{Pleinert2017}. We find that two-photon excess emission dominate single-photon excess emission into cavity mode for low pumping rates which implies that photon emission in cavity mode occurs mostly due to cooperative two-photon emission by QDs. Such cooperative emission occurs when correlation between QDs is established by photon exchange between QDs and cavity mode.
 Initially, on increasing pump strength single-photon, three-photon and four-photon excess emission into cavity mode changes from positive to negative and vice versa due to interference between different transitions corresponding to these processes.
 At higher pump strength single-photon excess emission dominates two-photon excess emission. For symmetric case, single-photon excess emission start dominating at higher pump strength than in the case of anti-symmetric case and the emission from individual QDs dominates. three-photon and four-photon excess emission  attain a slight positive value for moderate pumping strength in symmetric coupling case otherwise they show absorption. Eventually, with further increase in pumping strength excess photon emission into cavity mode is suppressed due to self quenching. In Fig. \ref{fig:Fig5}(b) $\&$ (d) the results corresponding to T=20K are plotted. On increasing temperature, single-photon emission into the cavity mode increased while the two-photon emission decreases and other multi-photon emissions are largely suppressed. Single-photon excess emission increases with rise in temperature only for the low pumping rates and suppressed at higher rates. The decline in excess emission in cavity mode is in agreement with the notion that with increase in phonon scattering processes results in dephasing and hinder cavity exciton coupling.

 In Fig. \ref{fig:Fig6} we present the results by varying the cavity detuning, $\Delta_{cp}$ for two different temperatures T=5K, 20K. We consider the typical value of coherent pump strength such that single-photon excess emission in cavity mode is zero, i.e. $\eta=3.0g_1$ for symmetric coupling in Fig. \ref{fig:Fig6}(a) $\&$ (b) and $\eta=1.9g_1$ for anti-symmetric coupling in Fig. \ref{fig:Fig6}(c) $\&$ (d). In Fig. \ref{fig:Fig6}, the two-photon excessive emission curve shows peak at $\Delta_{cp}=-\Omega'$, where the single-photon excessive emission has a dip. The cooperative effects between the QDs lead to predominant two-photon emission into cavity mode as mentioned earlier. Other higher-order photon excessive emissions are not significant in this off-resonantly coupled system and they show very small negative values implying multi-photon absorption from the cavity mode. In Fig. \ref{fig:Fig6} (b) $\&$ (d), with increase in temperature to 20K, single photon emission into cavity mode increased and two-photon emission decreased due to enhanced exciton-phonon scattering rates. Additionally, at $\Delta_{cp}=-\Omega'/2$ corresponding to the two-photon transition between pump dressed states of individual QDs, we can see small peaks in two-photon excess emission and four-photon excess emission. This aspect is also seen in mean photon number plots, Fig. \ref{fig:Fig4} (b) $\&$ (d). The appearance of this small peak is due to transitions induced by cavity photons between laser dressed states, where QDs are emitting two-photons independently or collectively emitting four-photons into the cavity mode. Therefore, it is appropriate to consider cavity detuning $\Delta_{cp}=-\Omega'$ and the pump rates where single-photon excess emission is negligible to realise coherently pumped cooperative two-QDs two-photon laser.
\section{\label{sec:level5}Conclusions}
To conclude, we have considered two quantum dots(QDs) coupled to a single mode photonic microcavity. The QDs are driven incoherently and coherently in strong coupling regime. We incorporated exciton-phonon coupling using polaron transformed master equation. We have derived laser rate equation and investigated single and multi-photon lasing in both incoherently and coherently pumped systems. We have explicitly calculated contribution from single and multi-photon excess emission into the cavity mode by exactly solving rate equation with high photon number truncation for convergence in the numerical results. In the case of incoherent pumping, resonantly coupled QDs with $\eta<2.0g_1$ show two-photon excess emission greater than single and other multi-photon emission into cavity mode. In the coherent pumping case, we have shown that cooperative effects lead to significant two-photon excess emission for cavity detuning $\Delta_{cp}=-\Omega'$. We have shown the behaviour of the system for symmetric($g_2=g_1$) and anti-symmetric($g_2=-g_1$) coupling to cavity mode. In both incoherent and coherent pumping cases, by selecting pump strength properly such that single-photon emission becomes negligible, the photons in the cavity mode are due to two-photon excess emission and the system behaves as two-photon laser.


\appendix
\section{COMPARISON OF SOME RESULTS OBTAINED FROM ME, SME $\&$ EQ.  \ref{eqn:meanphotonNumber}}\label{Appendix A}

\begin{figure}
    \centering
    \includegraphics[width=\columnwidth]{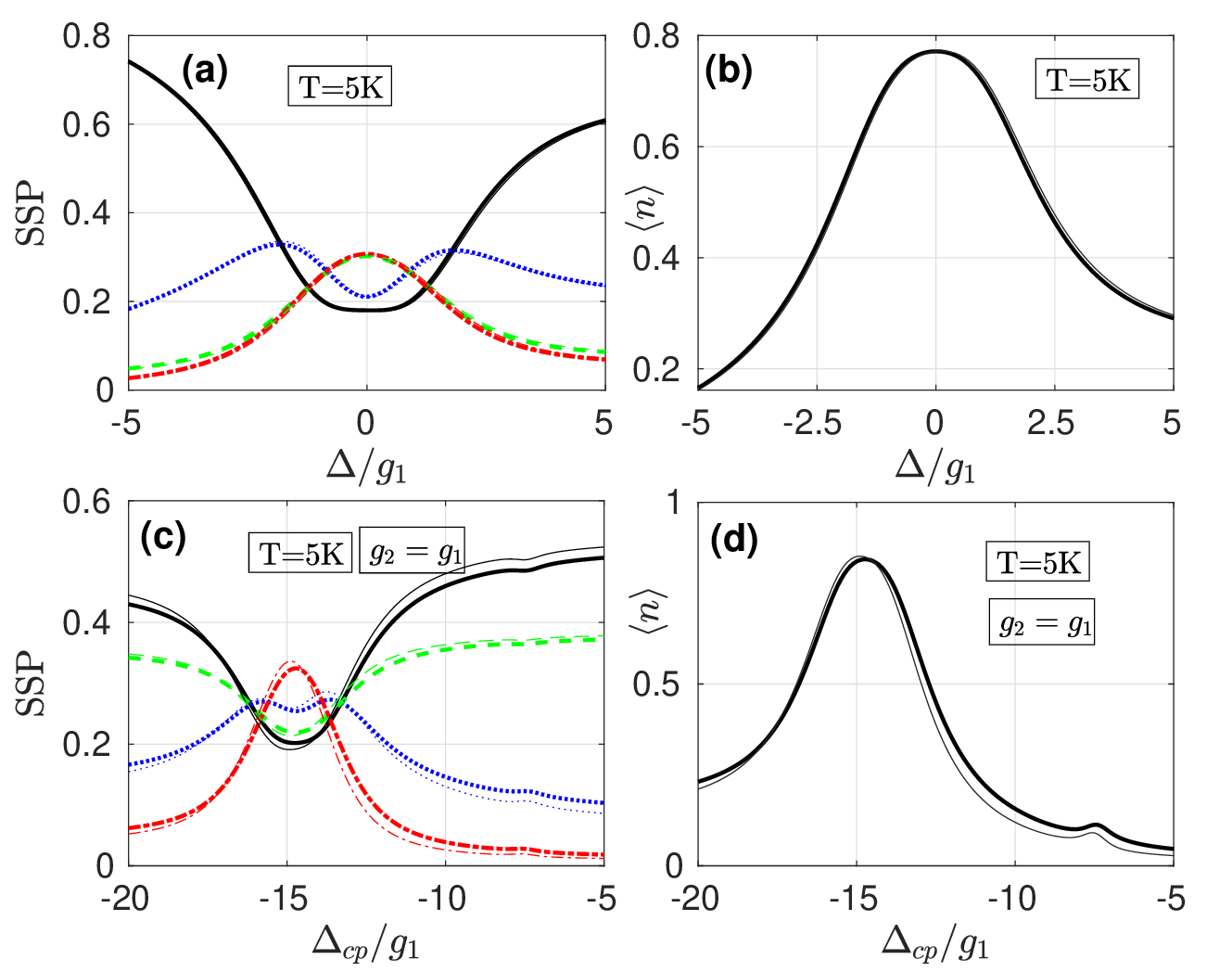}
    \caption{Comparison of steady-state populations of collective QD states calculated using exact master equation (thin) and simplified master equation (thick). In (a)$\&$(b) parameters are same as in Fig. \ref{fig:Fig2}(c) and in (c)$\&$(d) parameters are same as in Fig. \ref{fig:Fig4}(a). Colour scheme is same as Fig. \ref{fig:Fig2}}
    \label{fig:Fig7}
\end{figure}

\begin{figure}
    \centering
    \includegraphics[width=\columnwidth]{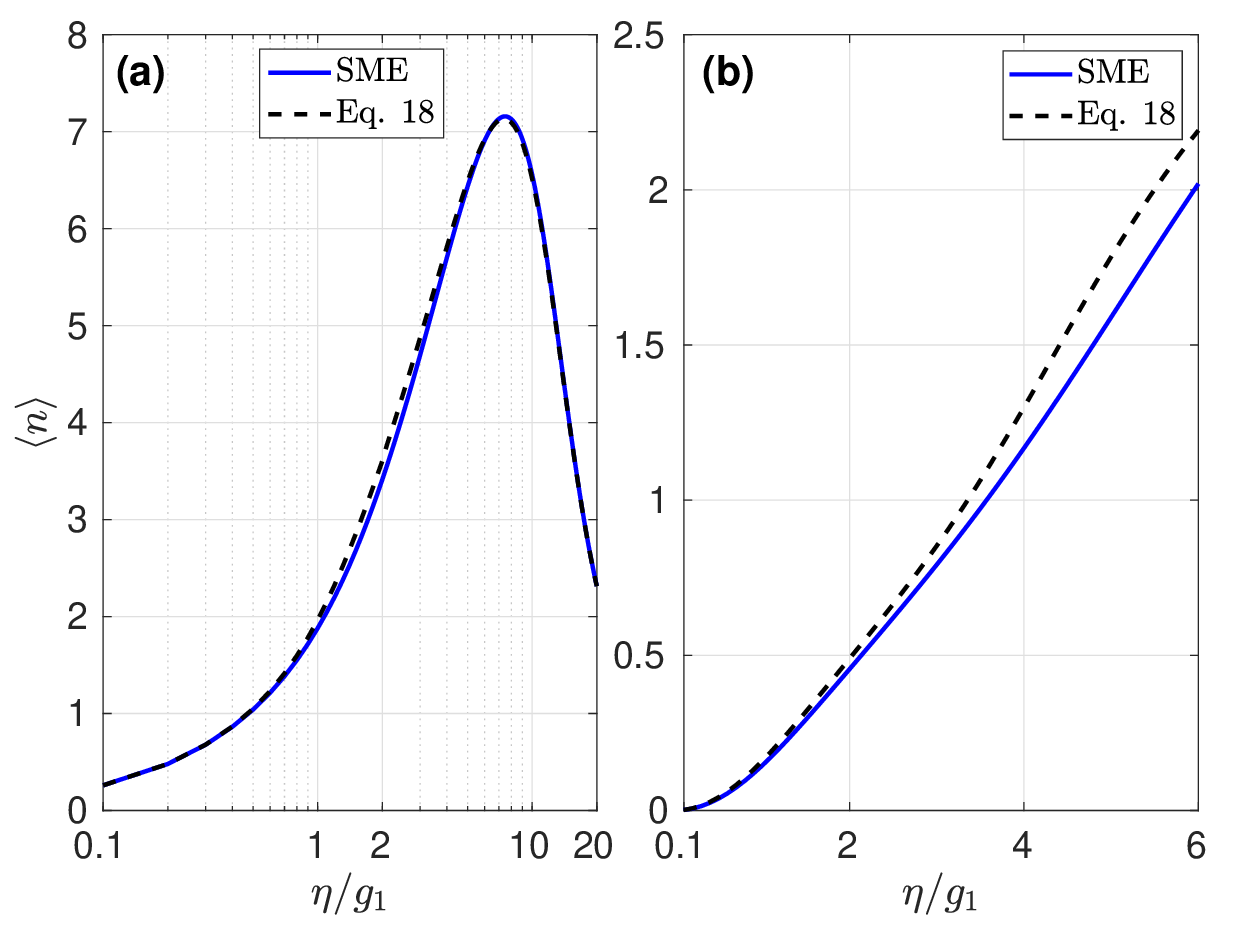}
    \caption{Comparison of mean photon number obtained from simplified master equation (solid blue) and the mean photon number obtained from laser rate equation \ref{eqn:meanphotonNumber} (dashed black) for a) incoherent pump and b) coherent pump, with variation in pump strength. The other parameters are same as in Fig. \ref{fig:Fig3}(a) for a) and Fig. \ref{fig:Fig5}(a).}
    \label{fig:Fig8}
\end{figure}

In Fig. \ref{fig:Fig7}, we compare the results of steady-state populations and mean photon number obtained from the exact master equation (\ref{eqn:incohME}) and the simplified master equations, (\ref{eqn:incohSME}),(\ref{eqn:cohSME}) for both incoherent and coherent pumping cases using same parameter considered in Fig.2 and Fig.4; respectively. By comparing the results of steady-state populations in collective QD states, $\ket{e_1,e_2}, \ket{+}, \ket{-}, \ket{g_1,g_2}$, and mean photon number in cavity mode it confirms the approximations made to obtain simplified master equations are very good to work in the range of parameters considered. 
 
We have also compared the mean photon number calculated from the simplified master equation for the incoherent pumping and the coherent pumping with the mean photon number equation (\ref{eqn:meanphotonNumber}) derived from the laser rate equation in Fig. \ref{fig:Fig8}. We can see the results are converging for the incoherent pump case exactly and are in good agreement for the coherent pump case. Here, we have considered terms up to $k=4$ in Eq. \ref{eqn:meanphotonNumber}. The results converge precisely if higher contributions are considered for coherent pump rates.

\bibliography{twoQDlasing}

\end{document}